\documentclass[namedreferences]{SolarPhysics}       
\usepackage[optionalrh]{spr-sola-addons} 
\usepackage{graphicx}               
\usepackage{color}                       
\usepackage{url}                         
\usepackage{lscape}
\usepackage{rotating}       
       
\newcommand{\solphys}{{\em Solar Phys.}}       

\usepackage{solaheader} 
\newcommand{\solareceiveddate}{9 Jan 2013} 
\newcommand{\solaaccepteddate}{31 July 2013}
\newcommand{\Xdoi}{s11207-013-0382-9}  

\begin{document}       
\begin{article}              
\begin{opening}
\title{Bimodal Distribution of Magnetic Fields and Areas of Sunspots}         
\author{Andrey G.~\surname{Tlatov}$^{1}$\sep       
Alexei A.~\surname{Pevtsov}$^{2}$}           
\runningauthor{Tlatov and Pevtsov}
\runningtitle{Bimodal Distribution of Magnetic Fields and Areas of Sunspots}
\institute{$^1$ Kislovodsk Solar Station of Pulkovo Observatory, PO Box 145,
Gagarina Str., 100, Kislovodsk, 357700 Russian Federation        
email: \url{tlatov@mail.ru}\\       
$^2$ National Solar Observatory, Sunspot, NM 88349, U.S.A.\\       
email: \url{apevtsov@nso.edu}}               
\begin{abstract}       
We applied automatic identification of sunspot umbrae and penumbrae
to daily observations from the {\it Helioseismic Magnetic Imager} (HMI) on
board the {\it Solar Dynamics Observatory} (SDO) to study their magnetic
flux ($B$) and area ($A$).  The results confirm a
previously known logarithmic relationship between the area of sunspots and
their maximum flux density.  In addition, we find that
the relation between average magnetic flux ($B_{\rm avg}$) and sunspot
area shows a bimodal distribution: for small sunspots and pores ($A 
\le 20$ millionth of solar hemisphere, MSH), $B_{\rm avg} \approx$
800 G (gauss), and for large sunspots ($A \ge$ 100 MSH), $B_{\rm avg}$ is
about 600 G. For intermediate sunspots, average flux
density linearly decreases from about 800 G to 600 G.  A similar
bimodal distribution was found in several other integral parameters of
sunspots. We show that this bimodality can be  
related to different stages of sunspot penumbra formation
and can be explained by the difference
in average inclination of magnetic fields at the periphery of 
small and large sunspots.

\end{abstract}                     
\end{opening}       
\section{Introduction}              
Sunspots are the longest studied activity phenomena on the Sun.  Recent
advances in high resolution observations and numerical modeling
have brought a better understanding of fine structure of individual sunspots,
their emergence, and evolution. Still, knowledge of statistical
properties of sunspots is important for understanding of their physical
characteristics as a distinct ``family'' of solar features, which in
turn could be invaluable for modeling of this solar phenomenon. Some of
the oldest records of sunspot statistical properties include their
coordinates and area ({\it e.g.}, \opencite{Hathaway2010}), with a later
addition of sunspot contrast and magnetic field strengths ({\it e.g.},
\opencite{Solanki2003}). All these previous data are based on
ground-based observations, which are subject to atmospheric image
distortion, interruption due to day-night cycle, and weather
conditions. The launch of {\it Solar Dynamics Observatory} (SDO) with the
{\it Helioseismic  Magnetic Imager} (HMI; \opencite{Scherrer2012}) on board
brings the studies of statistical properties of sunspots   to a new
level by providing nearly simultaneous observations of magnetic fields
and Doppler velocities with high spatial resolution and 
cadence. High (uniform) quality imaging observations from space 
also enable 
automatic identification of solar features. This work employs the HMI 
observations taken during 2010-2012 to study
statistical properties of sunspots such as their area and magnetic
field. To identify sunspots, we use an automatic method developed
by \inlinecite{Tlatov2012}. The data and their analysis are briefly 
described in Section 2. The magnetic field and area of sunspots 
(presented in Section 3) show a clear presence of two ``sub-families'' 
of  sunspots with distinct average properties. In Section 4 we discuss 
these findings.

\section{Data and Analysis Method} 
       
To identify sunspots, we employ the method developed by
\inlinecite{Tlatov2012}. Previously, the method was applied to
ground-based data from Kislovodsk Solar Station and space-borne
observations from SOHO/MDI. It showed a very good agreement with manual
and semi-automatic sunspot identifications. More recently, the method
was compared with STARA sunspot catalogue \cite{Watson2009}, and the
agreement was found to be excellent (A.~Tlatov, 2012, private
communication). The method was further modified to allow the
identification of multiple umbrae inside a common penumbra.

To determine the outer (quiet sun--penumbra) and inner
(penumbra--umbra) penumbral boundaries we employ  two known methods:
intensity threshold ({\it e.g.}, \opencite{Watson2009})  and the border
(gradient) method ({\it e.g.}, \opencite{Zharkova2005}). First,  a potential
sunspot is identified using an intensity threshold relative to a
limb-darkening function fitted to the data. For pixels below the
threshold, we apply a growing procedure; we add pixels to form
a closed boundary around the sunspot candidate if the intensity contrast
for  these pixels falls within the range of specified contrast. The
contrast  range for this step was determined by a trial-and-error
approach.  After the outer boundary of the sunspot is defined, we use the
distributions of intensity and contrast in pixels inside of that
closed boundary to  identify the penumbra--umbra boundary. Light bridges
are included in the total  area of the sunspot, but they are excluded from
umbral areas.  For additional details of this method, we refer the
reader to  \inlinecite{Tlatov2012}. For the analysis presented in this
article,  we correct sunspot areas for foreshortening.  

We use daily observations from SDO/HMI taken in
quasi-continuum (filenames hmi.Ic\_45s) during June 2010 -- September
2012. To minimize the size of data set and mitigate the effect 
of 
spacecraft's orbital motions, only one image per day (taken
at 5:00 UT) is used. The total number of features identified during the
above period is 14094 sunspots and pores. To minimize the projection
effects, we further limit the data set to features found within $30^\circ$
from solar disk center ($\mu = \sin \rho \le$ 0.5, where $\rho$ is 
angular heliocentric distance). 
Next, we classified all features into pores and
sunspots. Features for which the area of the umbra is 
equal to its total area are classified as pores and features with penumbra 
are classified as sunspots, respectively.  Thus, the final
data set consists of 
4648 features: 3027 pores and 1621 sunspots.

An error analysis of HMI longitudinal magnetograms was
published by  \inlinecite{Liu2012}. From this analysis, an upper limit
for random noise  in HMI longitudinal magnetograms is about 10 
Mx (maxwell) cm$^{-2}$ = 10 G (gauss) 
for 45-second  magnetograms (same type of magnetograms as used by
us). The noise increases  slightly as a function of heliocentric
distance. In addition, strong  fields measured in sunspots exhibit a 24-h 
modulation, which means that for higher
orbital velocities (relative to the Sun) the  calibration curve  for
longitudinal magnetograms becomes less accurate. To mitigate the
effect of orbital motions, we only select one magnetogram per day
taken at  about 5:00 UT. However, this approach does not completely
eliminate the effects of  spacecraft orbital motions.
On average, the amplitude of magnetic flux variations due to orbital
motions is about 2.7\% of the measured flux \cite{Liu2012}. An example in
\inlinecite{Liu2012} (see their Figure 5) shows residual variations
in the amplitude of field ranging from about 50 to 140 G (for the observed
longitudinal fields in the range of 1000--1200 G). These daily
variations in strong flux contribute to scatter in $B_{\rm MAX}$ shown  in
our Figure 1. In fact, after subtracting a fitted functional
dependence from data points shown in Figure 1, the residual scatter  is
about 120--140 G, which is in qualitative agreement with
\inlinecite{Liu2012}.  Such daily variations in magnetic flux will only 
affect $B_{\rm MAX}$, but not sunspot  areas.

After identifying sunspots and pores, their boundaries were superposed
onto the longitudinal magnetograms taken by HMI simultaneously with the
quasi-continuum images.  Magnetograms were used to compute the total
magnetic flux $\Phi$ and the maximum magnetic flux density 
($B_{\rm MAX}$).  In these calculations, we
corrected the observed magnetic field by $\cos \rho$.
Limiting the data set to $\rho \le$ 30$^\circ$ 
also helps in minimizing the effects of projection.  
Magnetic flux density is a product of field strength,  cosine of inclination
angle (relative to the line-of-sight) and the  fill factor (fraction of
a pixel occupied by magnetized plasma).  As the magnetic field is close
to vertical in sunspot umbrae, and the fill factor is close to unity
there, the observed longitudinal flux density corrected for $\cos \rho$
may serve as a good proxy for true field strength except for  areas of
very strong magnetic fields where longitudinal magnetograms may
exhibit magnetic saturation \cite{Hagyard1999}.
According to \inlinecite{Liu2012} SDO/HMI longitudinal
magnetograms  may saturate when the field strength exceeds 3200 G. 
During the 2010-2012 period included
in our study, very few sunspots were measured having their field strength
in excess of 3200 G (based on measurements of field strength by W. 
Livingston in Fe {\sc i} 1565 nm). \inlinecite{Pietarila2013}
found good correlation between longitudinal magnetograms from 
Vector Spectro-Magnetograph (VSM) on the {\it 
Solar Optical Investigation of the Sun}
(SOLIS) suite of instruments and SDO/HMI, although their correlation
varies across solar disk.

To further demonstrate that the longitudinal field provides a reasonable
proxy for true field strength in sunspot umbra, we use observations of
active region NOAA 7926 (18--25 November 1995) taken with the Advanced
Stokes  Polarimeter equipped on the Dunn Solar Telescope of the
National Solar Observatory, Sacramento Peak. 
Magnetic fields were derived via inversion of
Stokes profiles in the framework of Milne-Eddington model of stellar
atmospheres following \inlinecite{Skumanich1987}. In two
magnetograms taken three days apart, the maximum field strengths
derived by correcting the line-of-light flux by  the cosine of the
heliocentric distance $\rho$ were 1872~G and 1586~G. The
corresponding field strengths derived from full Stokes profile
inversion (taking into account the fill factor) were 1969~G and 1311~G, 
or within 5--20\%  of the field strength proxy derived from
the longitudinal magnetograms.

The correction using cosine of heliocentric angle works reasonably well for
umbral  magnetic fields which are more vertical. In the sunspot penumbra,
the magnetic fields become  more horizontal. Thus, the total and
average fluxes that we discuss in later sections refer to the vertical
flux.

\section{Relation between Magnetic Flux and Sunspot Area} 

Several studies ({\it e.g.}, \opencite{Houtgast1948}; \opencite{Rezaei2012})
found a correlation between the area of sunspots ($A$, in millionth of
solar hemisphere, MSH) and their recorded maximum field strength.  
\inlinecite{Ringnes1960} found the strongest
correlation between the logarithm of area and the field strength.

\begin{figure}[ht]
\includegraphics[width=1.0\textwidth,clip=]{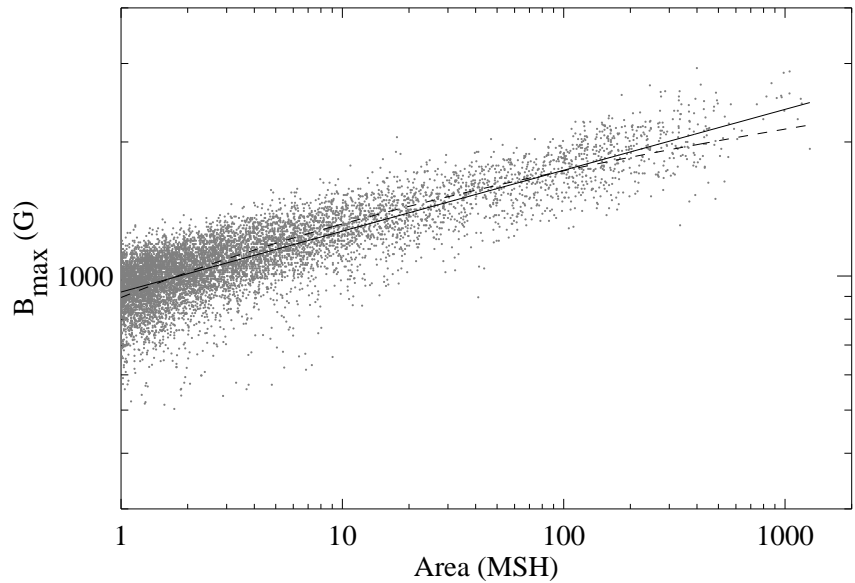}
\caption{Maximum field strength ($B_{\rm MAX}$) {\it vs.} area ($A$) 
of pores and sunspots. The black solid line shows the least-squares fit 
$\log B_{\rm MAX} =
(2.9640 \pm 0.0008) + (0.1367 \pm 0.0009) \times \log A$ to the data.
The dashed line corresponds to a fit 
$B_{\rm MAX} = (894.7640 \pm 2.1067) + (414.1450 \pm 2.458) \times \log A.$
}
\end{figure}

Figure 1 shows a scatter plot between maximum field strength ($B_{\rm MAX}$) 
in sunspot/pore and sunspot area ($A$). Two least-squares fits were 
made to the
data:  $\log B_{\rm MAX} = (2.9640 \pm 0.0008) + (0.1367 \pm 0.0009)
\times \log A$ (the solid line in Figure 1) and 
$B_{\rm MAX} = (894.7640 \pm 2.1067) + (414.1450 \pm 2.458) \times \log A$
(the dashed line). The $F$-test indicates that both of these models
fit the data equally well. Pearson's correlation coefficient 
$r_{\rm P}$=0.8731
is slightly higher for $\log A - {\rm linear}~ B_{\rm MAX}$
as compared with the $r_{\rm P}$=0.8410 for $\log A - \log B_{\rm MAX}$
relationship. However, visually, it appears that $\log A
- {\rm linear}~ B_{\rm MAX}$ fit works well only for the range of smaller
areas and lower field strengths. The $\log A - \log B_{\rm MAX}$ fit
works well for all ranges in area and field strength.
\inlinecite{Pevtsov2013} noted possible correlation between the steepness
of the $\log A - \log B_{\rm MAX}$ relation and the amplitude of the next 
solar cycle. The solar-cycle dependence of the 
$\log A$ - {\rm linear} $B_{\rm MAX}$
relationship was described by \inlinecite{Ringnes1960}, 
who found the following scaling coefficients; for 1917--1924 (cycle 15): 980,
1924--1934 (cycle 16): 970, 1934--1944 (cycle 17):1410, 1945--1954 (cycle 18):1610,
and 1954--1956 (rise of cycle 19):1710.
The amplitudes of sunspot cycles progressively increase from cycle 15 to
cycle 19.  If the scaling coefficient derived from the fitted relationship 
in Figure 1
follows the above trend, solar cycle 24 is expected to be
much lower in amplitude than cycles 15 and 16, which is in agreement
with the current predictions for cycle 24. 

While there is no physical model explaining the observed
relationship between the magnetic field and area of sunspots, $\log A -
\log B_{\rm MAX}$ relationship can be derived from the distribution
of the magnetic field of a dipole situated at a certain depth below
the photosphere \cite{Ikhsanov1968}. The dipole field was found to be a
good representation of sunspot magnetic field
\cite{Bumba1960,Skumanich1992}.

\begin{figure}[ht]
\includegraphics[width=1.0\textwidth,clip=]{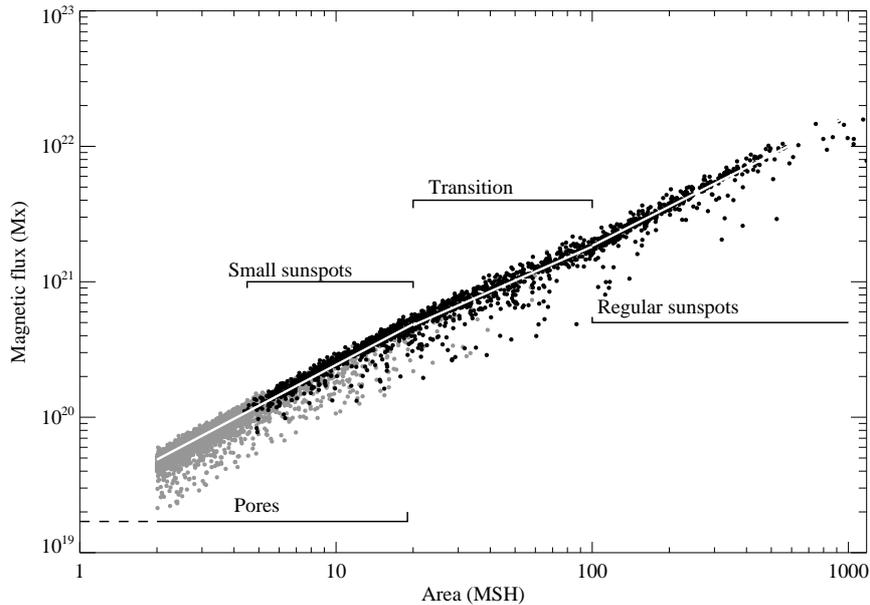}
\caption{Total magnetic flux as a function of the area of pores (gray) 
and sunspots (black). Because pores and small sunspots partially 
overlap, horizontal bars mark approximate range in their areas. 
Also marked are ranges for regular sunspots and transition between 
small and regular sunspots. Three straight (white) lines plotted 
over the data points correspond to linear fits to the data for areas 
smaller than 20 MSH (pores and small sunspots), between 20 and 100 MSH
(transition), and larger than 100 MSH (regular sunspots).
}
\end{figure}

Figure 2 shows a scatter plot between total magnetic flux ($\Phi$) and 
sunspot/pore area ($A$). As the location of pores and small sunspots partially
overlaps in this plot, their approximate ranges in size are identified by
horizontal bars. The smallest sunspots in our data set have an area of about
4.5 MSH, which corresponds to a circular feature of about 4.2 Mm in
diameter.  The largest pores have an area of about 19 MSH, which
corresponds to a circular feature of approximately 8.5 Mm in diameter.
For comparison, the smallest sunspots reported by \inlinecite{Bray1964}
were about 3.5 Mm in diameter, and the largest pores were observed to
be of 7 Mm in diameter. In the following
discussion, we refer to sunspots with $A <$ 20 MSH as small sunspots and
to those with $A >$ 100 MSH as regular sunspots. The area between 20 and 100
MSH is referred to as the transition sunspots.  
The data shown in Figure 2 include 3027 pores,
624 small sunspots, 560 transition sunspots, and 437 regular sunspots.
The selection of ranges for separating small, transition, and regular 
sunspots is based on the transitions between these sunspots clearly 
identifiable in Figure 2.

Figure 2 shows one peculiar feature that, perhaps, deserves  a
separate explanation. The scatter in flux (in the $y$-direction) is
asymmetric: it only extends towards the lower values, while
there appears  to be an upper limit in it. We attribute
this to a residual effect of  the spacecraft orbital  motions. As
\inlinecite{Liu2012} showed, the magnetic flux  measured in a sunspot
varies with 24-h periodicity. In this variation,  the measurements
taken on the same sunspot during a single day will be equal to or
systematically lower than the ``true'' flux of that sunspot. At the same
time,  sunspot areas will not be affected by orbital motions. Thus,
the upper limit  in magnetic flux in Figure 2 represents the ``true'' value
of magnetic flux for measured features, and asymmetric scatter extending below
that ``true'' value is due to systematic  variations associated with
orbital motions.  A similar upper limit can also be seen in  Figure 1,
although it is much less clear in comparison with Figure 2.  

Although Figure 2 looks very similar to Figure 1, there is one
important difference: the slope is different for areas smaller than
about 20 MSH and for those larger than 100 MSH.  A linear
fit to the area--flux relationship in Figure 2 shows a clear change between
small and transition sunspots, and between transition and regular sunspots 
(compare inclination of
three linear segments fitted to  the data in Figure 2). If one has to
fit a continuous function to the data in Figure 2, it appears that an $A
\sin(\log A)$ function fits reasonably well. 
The curve corresponding
to this function will overlap with three linear segments shown in 
Figure 2, and thus, is not shown in Figure 2.

The ranges for ``small'' and ``regular'' sunspots are not exact  and thus, 
they do have some small degree of uncertainty.
This uncertainty, however, has no effect on our conclusions.

The transition from one
slope to the other is more profound in the relationship between area $A$
and average magnetic flux $B_{\rm avg}$ (Figure 3).  For pores,
$B_{\rm avg}$ slightly increases with area, albeit with significant
scatter in the data.  For small sunspots, the average magnetic
flux stays constant at approximately 850 G, and for regular sunspots 
$B_{\rm avg} \approx$ 600 G.  For transition sunspots, $B_{\rm avg}$ decreases
with sunspot area almost linearly.

\begin{figure}[ht]
\includegraphics[width=1.0\textwidth,clip=]{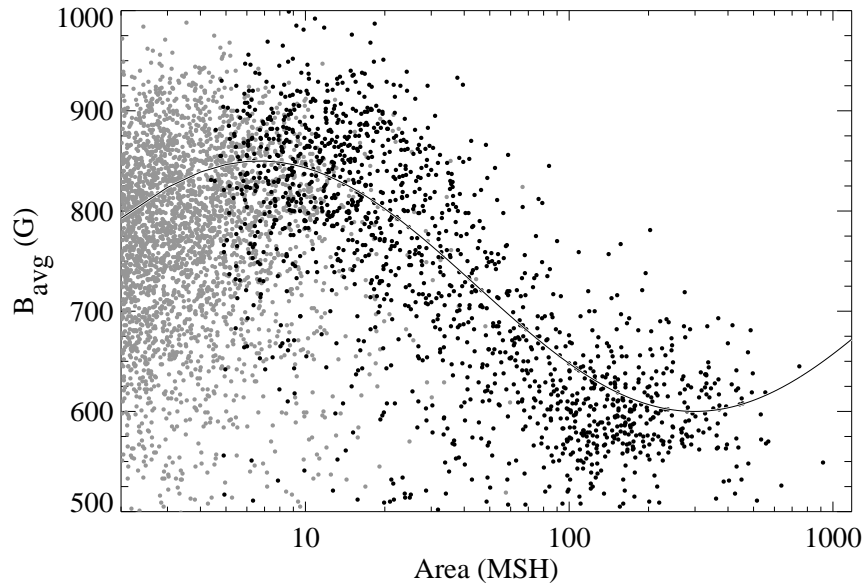}
\caption{Average magnetic flux {\it vs.} area of pores (gray dots) 
and sunspots (black dots). The thin solid line corresponds to a 
$\sin(\log A)$ function.}
\end{figure}

\begin{figure}[ht]
\includegraphics[width=1.0\textwidth,clip=]{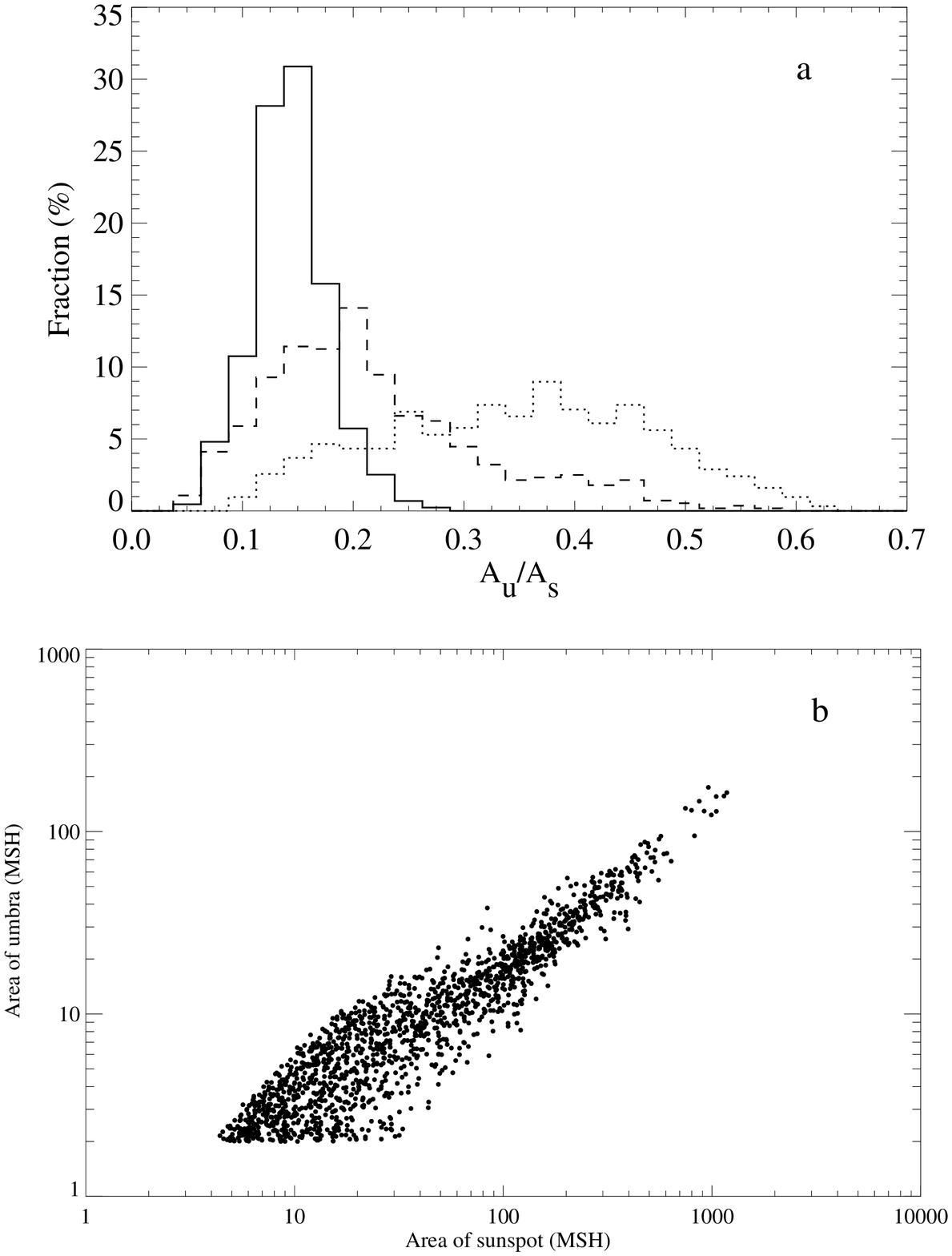}
\caption{(a) Normalized distributions of the ratio of umbral area 
($A_{\rm u}$)
to total sunspot area ($A_{\rm s}$) for regular sunspots (solid line, 
$A_{\rm s} \ge$100 MSH), 
transition sunspots (dashed, 20 MSH $< A_{\rm s} <$100 MSH) and
rudimentary (small) sunspots (dotted, $A_{\rm s} \le$ 20 MSH).
(b) The area of umbra ($A_{\rm u}$) {\it vs.} the total area of sunspot 
($A_{\rm s}$).}
\end{figure}

\section{Bimodal Distribution of Sunspots}

Geometric properties of ``small'' and ``regular'' sunspots form two
distinct groups. Figure 4a shows the distributions of the ratio  
$A_{\rm u}/A_{\rm s}$ between the area of sunspot umbra ($A_{\rm u}$) 
and the total area of sunspot ($A_{\rm s}$) for
three groups. For regular sunspots, the distribution is very narrow,
and it peaks at about $A_{\rm u}/A_{\rm s}$=0.156$\pm$0.034. This is
in agreement with previous studies, {\it e.g.}, $A_{\rm u}/A_{\rm
s}$=0.168--175 in \inlinecite{Antalova1991} and references therein (the
above values were derived from the ratio of the radii of umbra and sunspot
listed in Table 1 of \opencite{Antalova1991}).  Some of these
previous studies suggest that the umbra/sunspot area ratio changes
with sunspot field strength \cite{Antalova1991}. On the other hand, no
variation in $A_{\rm u}/A_{\rm s}$ with solar cycle was found by
\inlinecite{Beck1993}.

For small sunspots, the distribution of $A_{\rm u}/A_{\rm s}$ ratio
is very broad with the average of 0.359$\pm$0.119.  For transition
sunspots, $A_{\rm u}/A_{\rm s}$=0.231$\pm$0.099 (see Figure 4a).
This can also be seen in Figure 4b, which shows the relationship between 
the area of umbra and area of sunspot.

A narrow width of the distribution of $A_{\rm u}/A_{\rm s}$ for regular
sunspots suggests lack of dependence of the area ratio on the field strength
for these sunspots.

The bimodal distribution is present in several other parameters. For
example, the relationship between $B_{\rm MAX}$ in sunspots and
average magnetic flux density in penumbra ($B_{\rm avg,p}$) shows two 
distinct tendencies: for
smaller sunspots with $A_{\rm s} \le$ 50 MSH, $B_{\rm avg,p}$ increases
with $B_{\rm MAX}$. For larger sunspots with $A_{\rm s} > $ 50 MSH,
$B_{\rm avg,p}$ stays nearly constant at about 500 G, independent of
$B_{\rm MAX}$. The ratio of areas $A_{\rm u}/A_{\rm s}$ {\it vs.} the ratio of
fluxes in the umbra ($\Phi_{\rm u}$) and in the entire sunspot 
($\Phi_{\rm s}$) also
forms two distinct relationships for smaller and larger sunspots. Due to
space limitations, these  other relationships are not shown as graphs.

\section{Discussion}

Our analysis of properties of sunspots automatically selected in
SDO/HMI daily images confirms the logarithmic relationship between
maximum magnetic field strength in sunspots and their total area. Both
$\log - \log$ and $\log -$ linear functional relationships fit the
data relatively well, although $\log - \log$ least-squares fit appears
to work better. Although several studies had established a strong
correlation between the area of sunspots and their maximum field strength,
the nature of this relationship is not well understood. In
addition, the measurements of (maximum) magnetic fields may differ
between instruments, while the measurements of areas are more
instrument independent.  For example, scattered light and strong
blending in the Ni {\sc I}  676.8 nm line in the umbrae of sunspots
could be a factor for $B_{\rm MAX}$ measurements for some instruments
({\it e.g.}, MDI). These factors could result in a different $\log A - \log
B_{\rm MAX}$ relationship as found by us.

We also find a bimodal distribution of properties of
sunspots, depending on their size.  Small sunspots and pores follow the
same relationship between the magnetic flux and area. We see a
significant overlap in size and flux between these two solar features.
This suggests that small sunspots may originate from pores, and not
develop as independent features. The scenario that sunspots evolve from
pores was previously proposed by
several researchers \cite{Rucklidge1995,Skumanich1999}.  In particular,
\inlinecite{Rucklidge1995} suggested that when a pore reaches a
critical size, it will abruptly make a transition to a sunspot. Indeed, 
our data
indicate a transition between small (rudimentary) sunspots and regular
sunspots. However, the shape of transition is rather gradual, which is
different from a hysteresis-type transition as in
\inlinecite{Rucklidge1995} model.  Interestingly, functional relationship
between sunspot radius and magnetic flux based on the $\sin(\log A)$
function fitted to Figure 3 is similar to the one shown in Figure 4 of
\inlinecite{Skumanich1999}. By itself, the choice of $\sin(\log A)$
function was made arbitrary based on a visual fit to the data in Figure
3. While we have no physics-based explanation for this functional
form, we note that it represents the $\log A - \log \Phi$ relationship in 
Figure 2 quite well. Although not shown in Figure 2, the
umbrae of sunspots follow the same trend as small sunspots and pores.
For umbrae, the trend continues through the area occupied by ``transition''
sunspots, but unlike sunspots, the umbrae do not change  
the slope of $\log A - \log \Phi$ relationship.

The transition between small and regular sunspots may be associated with
the formation of regular penumbra. When the sunspot penumbra
starts forming, the magnetic field becomes more horizontal, and thus,
the relation between the vertical flux and the area of sunspot changes. This
change corresponds to transition sunspots in Figure 2. For sunspots
with fully developed penumbra, the ratio between the areas of umbra and 
sunspot as
a whole remains constant even though the sunspots may continue to grow
in size.  These changes in flux--area and umbra--sunspot area
correlations are in a qualitative agreement with the scenarios of sunspot
formation described, {\it e.g.}, by  \inlinecite{Zwaan1992} and
\inlinecite{Rezaei2012}.  \inlinecite{Collados1994} studied three
sunspots with well-developed penumbra, and found  systematic
differences in sunspots of different size. They concluded that
sunspots with small and large umbrae may require different models  to
represent their atmospheres.

The longitudinal field is a projection of full vector field to 
the line-of-sight direction. Thus, it is possible to derive some 
information about the inclination of vector field to solar surface from
the line-of-sight data, albeit with some restrictive assumptions.
For example, one can assume that the magnetic field in a sunspot does not 
change significantly over several days as it travels across the 
solar disk. In the framework of that simple model, the line-of-sight field 
will show a different evolution depending on the inclination of vector field
to solar surface. Below we exploit this approach to demonstrate that the 
inclination of magnetic field at the periphery of sunspots is
different for sunspots of different size.

Figure 5 shows the ratio of the magnetic field
in the disk center side of the penumbra at the 
distance of 2/3 of sunspot radius from
the center of sunspot ($B_{2/3}$) to the magnetic field at the center of
sunspot ($B_{\rm c}$) as a function of heliocentric distance $\mu =
r/R$. (Here we compare only magnetic fields in the disk center side of 
penumbra and at sunspot center to avoid potential issues with 
polarity reversal of magnetic field at the periphery of sunspots due to 
projection effects).
For this plot, we include all identified sunspots up to $\mu$ =
0.98. As a reminder, the data discussed earlier were
limited to $\mu \le$ 0.5. Next, we computed median values of
$B_{2/3}/B_{\rm c}$ within $\pm$0.1 intervals centered at $\mu =
$0.1, 0.3, 0.5, 0.7, and 0.9.  For small sunspots (triangles in Figure 5), 
$A_{\rm s} \le$ 20 MSH), the ratio $B_{2/3}/B_{\rm c}$
is about 0.9,  and it does not vary
significantly with the heliocentric distance.  This indicates that on
average, the magnetic fields are nearly vertical in smallest sunspots.
Transition sunspots with areas 20 MSH $< A <$ 100 MSH show a
small variation in $B_{2/3}/B_{\rm c}$ with $\mu$. And for
regular sunspots, this variation with the heliocentric distance is much
stronger. Since we use the line-of-sight (LOS) magnetograms, $B_{\rm
c}$ will vary with heliocentric angle as $B^{\rm LOS}_{\rm c} = B_{\rm c}
\cos \rho$, where $\rho$ is angular heliocentric distance 
(this assumes that the magnetic field is vertical at the center of 
the sunspot). For a
magnetic field inclined from the vertical direction by an angle $\gamma$, 
the LOS component can be expressed as $B^{\rm LOS} = B \cos (\gamma - \rho)$.
Therefore, $${{B^{\rm LOS}_{2/3}} \over {B^{\rm LOS}_{\rm c}}} = 
{{B_{2/3} \cdot \cos(\gamma - \rho)} \over {B_{\rm c} \cdot \cos \rho}} = 
{B_{\rm 2/3}
\over {B_{\rm c}}} (\cos \gamma + \sin \gamma \cdot \tan \rho).$$
The least-squares fit to the data by this function returns progressively
larger inclination angles in sunspot penumbra (at the distance of 
$\frac{2}{3} R_{\rm s}$ from sunspot center):  $\approx$ 1$^\circ$ 
(small sunspots),
15$^\circ$ (transition sunspots), and 38$^\circ$ (regular sunspots).

\begin{figure}[ht]
\includegraphics[width=1.0\textwidth,clip=]{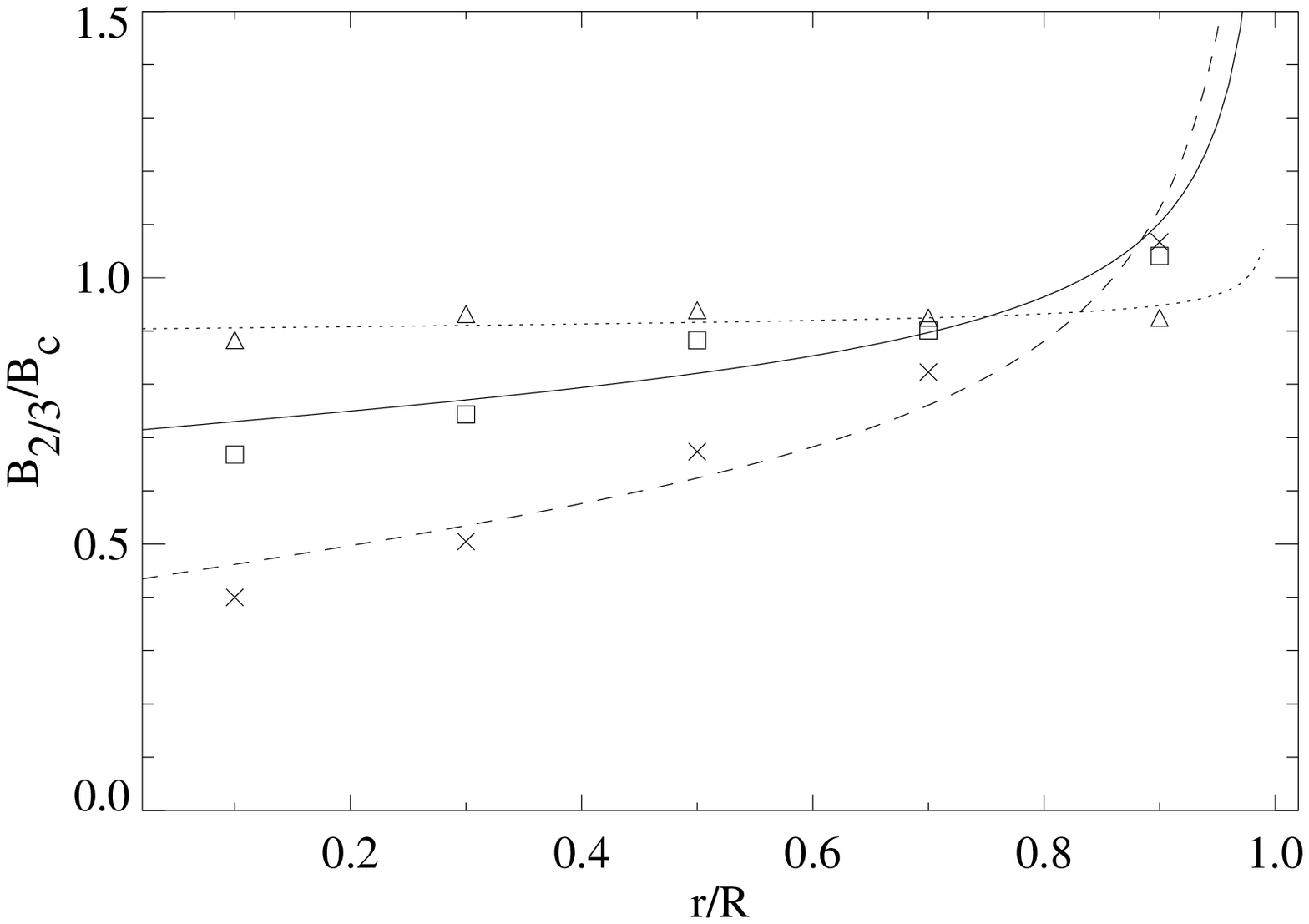}
\caption{The ratio $B_{2/3}/B_{\rm c}$ of the magnetic field at the distance 
of 2/3 of the sunspot radius to the magnetic field at sunspot center as a
function of
heliocentric distance $r/R$ for small sunspots (triangles and
dotted line), transition sunspots (squares and solid line), and regular
sunspots (crosses and dashed line). Symbols indicate median values
computed for five intervals in $r/R$ (in 0.2 step).
}
\end{figure}

The difference in average inclination of penumbral magnetic field (at
$\frac{2}{3} R_{\rm s}$) between small, intermediate, and regular sunspots 
supports our early suggestion
that the bimodal distribution of properties of sunspots may be associated
with different stages of sunspot formation. Small sunspots may have
rudimentary and irregular penumbra, which has not yet evolved to a regular
penumbra with more horizontal fields as in regular sunspots.  In
transition sunspots ($A$ between 20 and 100 MSH), the average inclination
of magnetic fields in the penumbra increases, and rapidly, fields become more
horizontal.  

The change in the inclination of vector field for sunspots at different 
stages of
its development was observed using data from the Advanced Stokes 
Polarimeter (using magnetograms of active region NOAA 7926
described in Section 2). The main sunspot in this active region
was decaying over the period of observations. On the day when the sunspot had
a well-developed penumbra, the average inclination of vector magnetic field 
in the penumbra (at about 2/3 of the sunspot radius) was about 
25-30$^\circ$, while in the umbra the field was nearly vertical 
(the inclination angle of about 80$^\circ$).
Three days later, when this sunspot lost most of its penumbra, 
the inclination of magnetic field in the umbra did not change, but the field
at 2/3 of radius of the sunspot became more vertical (the inclination 
increased to 55-60$^\circ$). While this is a single example, it supports our 
inferences about difference in inclination of magnetic field in sunspots
with different stages of penumbral development.

\inlinecite{Vitinsky1986} (see also, \opencite{Dmitrieva1968}) reported 
that the distribution of
areas of sunspots shows two distinct peaks.  One maximum in that
distribution corresponded to areas of about 8-- 13 MSH.
\inlinecite{Vitinsky1986} associated this maximum with the location of
small sunspots often developing in inter supergranular spaces
(vertices), where the boundaries of three supergranules meet together.
The second maximum in the size distribution was found at $A$= 100-150
MSH, which corresponds to an approximate size of supergranules.
Recently, \inlinecite{Nagovitsyn2012} reported similar bimodal
distribution of sunspot areas. Following the arguments given by 
\inlinecite{Vitinsky1986}, for small
sunspots located at the vertices of supergranular cells, there is an
additional force from converging supergranular flows (as well as a
strong downflow at the vertex location), which may result in
increased average magnetic field in small sunspots and pores. When
sunspots grow and get comparable in size with a supergranule, this
additional force decreases. 
These early ideas need to be further verified with more recent modeling 
and observations.
Finally, the results presented in this article are derived
using longitudinal magnetograms.  Additional studies based on high
spatial resolution vector magnetograms may further improve the
understanding of the bimodal distribution of sunspots reported here.

\begin{acks}              
A.G.T. acknowledges support by the Russian Foundation for Basic
Research (RFBR) and the Russian Academy of Sciences (RAS).  The
National Solar Observatory is operated by the Association of
Universities for Research in Astronomy (AURA), Inc. under a
cooperative agreement with the National Science Foundation.

\end{acks}

\end{article}                   
\end{document}